# Design Parameters and Commissioning of Vertical Inserts Used for Testing the XFEL Superconducting Cavities

J. Schaffran[a], Y. Bozhko[a], B. Petersen[a], D. Meissner[a], M. Chorowski[b] and J. Polinski[b]

[a]*Deutsches Elektronen Synchrotron DESY, Notkestr. 85, D-22607 Hamburg, Germany*
[b]*Wroclaw University of Technology, Wyb. Wyspianskiego 27, 50-370 Wroclaw, Poland*

**Abstract.** The European XFEL is a new research facility currently under construction at DESY in the Hamburg area in Germany. From 2015 on, it will generate extremely intense X-ray flashes that will be used by researchers from all over the world. The superconducting XFEL linear accelerator consists of 100 accelerator modules with more than 800 RF-cavities inside. The accelerator modules, superconducting magnets and cavities will be tested in the accelerator module test facility (AMTF). This paper gives an overview of the design parameters and the commissioning of the vertical insert, used in two cryostats (XATC) of the AMTF-hall. The Insert serves as a holder for 4 nine-cell cavities. This gives the possibility to cool down 4 cavities to 2K in parallel and, consequently, to reduce the testing time. The following RF measurement, selected as quality check, will be done separately for each cavity. Afterwards the cavities will be warmed up again and will be sent to the accelerator module assembly.

## INTRODUCTION

Six vertical Inserts are used as support structures for XFEL-nine-cell cavities during a cold test in a cryostat at 2K, to qualify all cavities for the operation in the XFEL accelerator. Each Insert could be equipped with up to four cavities. The RF measurements, selected as quality check, will be done separately for each cavity. To fulfill all constrains the design parameters have to be chosen carefully. The purpose of this paper is to present the design parameters in due consideration of the mechanical, cryogenic, particle free ultra-high vacuum and RF requirements as well as the commissioning and qualification of the Inserts. The temperature distribution on the support structure during the first cool down is discussed, too, just like its influence on the cool down rate.

## DESCRIPTION

The Insert can be divided into two parts, the upper one and the lower one. A picture of the complete Insert equipped with four cavities is given in figure 1. The upper part, designed by Deutsches Elektronen Synchrotron DESY and manufactured at Wroclaw University of Technology, acts as support structure of the lower part. It mainly consists of the top plate with all sub flanges (used for the RF-, vacuum and cryogenic system), four thermal shields and the supporting rods. The top plate was built under considerations of the European Pressure vessel directive and is demonstrated in figure 2. The lower part, acting as cavity support structure for four cavities, was designed and manufactured by Deutsches Elektronen Synchrotron DESY. Each cavity is connected to the vacuum and RF-system separately, so that each cavity could be operated as an own system.

### Mechanical Constrains

The Insert was designed for supporting four cavities equipped with or without helium containment. The overall dimensions are determined by the dimensions of the vertical test cryostats XATC1 and XATC2 and given as 3700mm in the depth and 1050mm in the diameter including the cavities with its accessories and a safety margin of 50mm. To avoid a deformation of the cavities, the flatness of the two cavity holding support plates has to be less than 0.3mm against each other. The thickness of the plates and the supporting structure is dimensioned by the forces of the cavity, dominantly the vacuum force of 7kN performed on each cavity. The axial orientation of the cavities is devoted to the connection of the vacuum pipes and the RF system. A strainless installation of the cavities is mandatory, to avoid a deformation resulting in a degradation in performance. This is guaranteed by a defined and carefully performed installation procedure realized by especially designed connection flanges with variable length. An anti-friction bush, usable for cavities equipped with He-pressure tank only, prevents a deformation of the cavity during the cool down process. An overview of the lower part is given in figure 3. The permeability of all materials used is not allowed to exceed a value of 1.1 at 1000Oe.

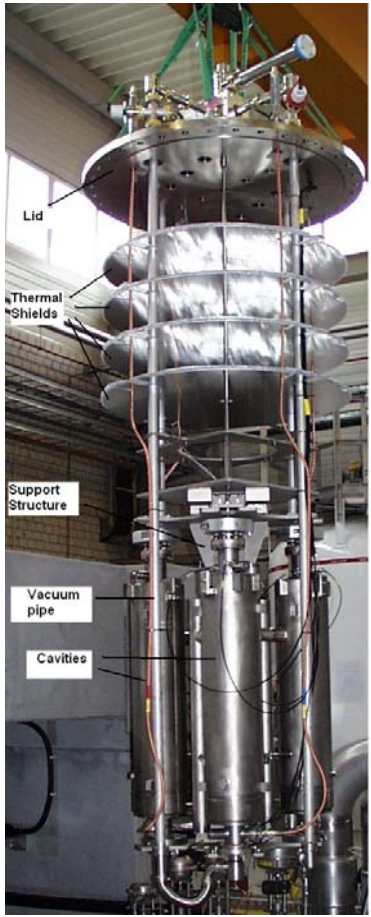

**FIGURE 1.** The Vertical Insert of the XATC-Cryostat

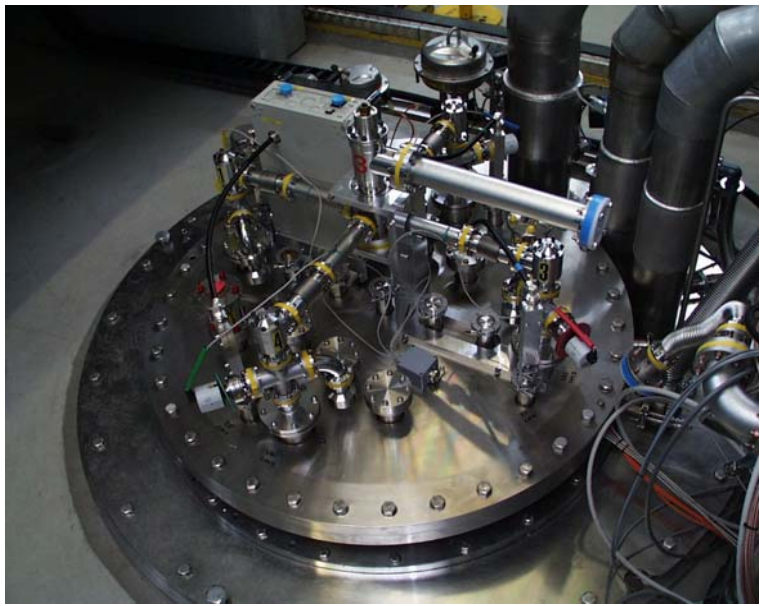

**FIGURE 2.** The Lid of the XATC-Cryostat

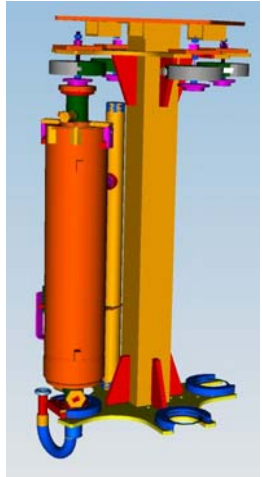

**FIGURE 3.** The Support Structure of the vertical Insert equipped with one Cavity

## Cryogenic System

The cavities have to be operated at a temperature of approximately 2K. The shape stability of the cavity is very important. If the cavities are stressed by mechanical forces a detuning and therefore degradation in their performance is possible. To avoid this, the main support structure consisting of a rectangular pipe was made out of Titanium (Type II). This material was also used for the helium containment of the cavities and has a comparable thermal expansion coefficient as the cavity itself. The thermal expansion was considered by pre-tuning of each cavity at room temperature.

Four thermal shields, installed in a length of 1m, and an optimized support structure were taken into account for reducing the heat load.

Eight temperature sensors, one level gauge and one heater were installed providing the cryogenic operation and commissioning.

## Vacuum System

The design parameters of the vacuum systems are determined by the length of the vacuum pipes and the requirements during the cavity installation process and normal operation. Especially, an oil- and particle free condition of all vacuum parts are mandatory. An overview of the complete system is given in figure 4. DN40-pipes provide a sufficient pumping speed on the cavities [1]. To allow a transportation and installation of pumped cavities, each one is equipped with a full metal angle valve. It has to be ensured that all this four valves are opened before installation into the cryostat starts. Four more valves installed on the top plate allow the operation of the cavities separately at cryogenic operating conditions and to prevent a warm-up, if cavities are leaky. Four bursting discs prevent a deformation of the cavity during the warm-up process in case of helium contamination due to a leak. Five vacuum gauges allow the observation of the vacuum at various operation conditions. A mass spectrometer installed on the pumping unit allows a mass spectrum analysis of all cavities, separately or commonly.

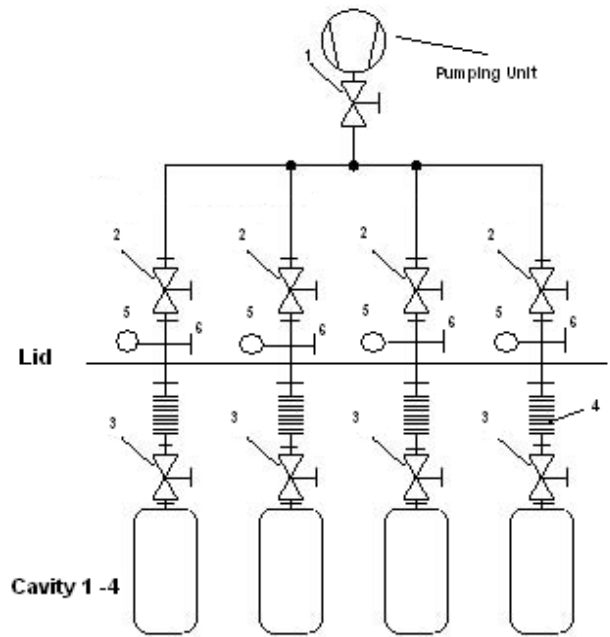

**FIGURE 4.** The vacuum system of the vertical Insert of XATC (1-Main valve, 2-Vacuum valve at room temperature, 3-Full metal angle valve, 4-Bellows, 5-Vacuum gauge, 6-Berst disc)

### RF System

Each cavity is equipped with an own RF-System consisting of High-Q antenna, pick-up antenna and 2 HOM-antennas to allow the measurement of the gradient and frequency spectrum for the quality control one by one. To prevent the propagation of neutron radiation during the RF-tests, four radiation shields are installed above the cavities. A detailed description of the RF-system is given in [2]

## COMMISSIONING AND QUALIFICATION

The commissioning of the vertical Inserts was divided into two parts. The first part was a mechanical check of compatibility of components, especially the installation process of the cavities. The mechanical installation of four cavities equipped with He-pressure tank or not, was tested successfully. The cavities could be installed without deformation, which could be observed by a spectrum measurement (hereby deformations in the range of a few μm can be detected by measuring the frequencies of the fundamental mode at room temperature). A mass spectrum analysis of each cavity demonstrated an excellent vacuum connection without contamination. The connection of all vacuum components as the RF-components could be done without problems, too. At least the installation into the cryostat was done easily.

The second part was the cool down of the Insert to 2K. Four cavities were installed to the Insert, cooled down to 2K, where a further spectrum measurement (Frequencies of fundamental modes, lock on $9/9\pi$ mode, phase [$P_{for}$, $P_{trans}$]) and a power rising (qe curves: max. $E_{acc}$, $Q_0$-factor vs. $E_{acc}$, Radiation vs. $E_{acc}$) were performed. After warming up again an additional spectrum measurement confirmed the cavity shape. Eight temperature sensors give information about the temperature gradients on the support structure. This gives the possibility to define the cool-down and warm-up rate. The positions of all sensors are shown in figure 5. All four cavities have been measured in a qualified cryostat before so that its performance is well-known. The level gauge and the heaters, installed on the lower part of the insert, were tested as were their integration in the cool-down process.

The cool-down process was controlled by the temperature difference on the supply and return line. The maximum allowed difference for the first test run was 50K. The temperature profile during the cool-down is given in figure 6a and 6b. During the complete cool-down process the temperature gradient in horizontal direction is less than 4K, in vertical direction less than 35K. An unexpected deformation on the cavities could not be detected at operation conditions. The heat load of the complete cryostat – liquid helium vessel and insert - was calculated to 7W [3].

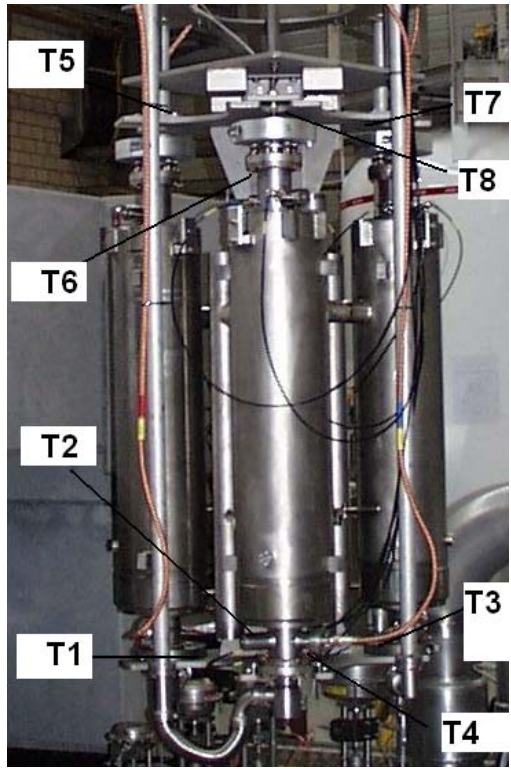

**FIGURE 5.** Positioning of the Temperature Sensors on the lower Part of the Insert

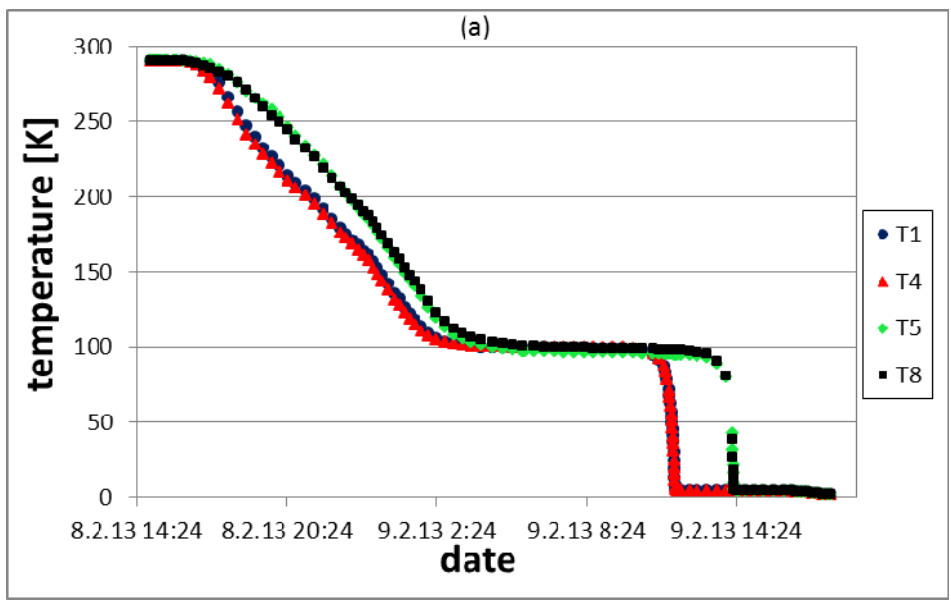

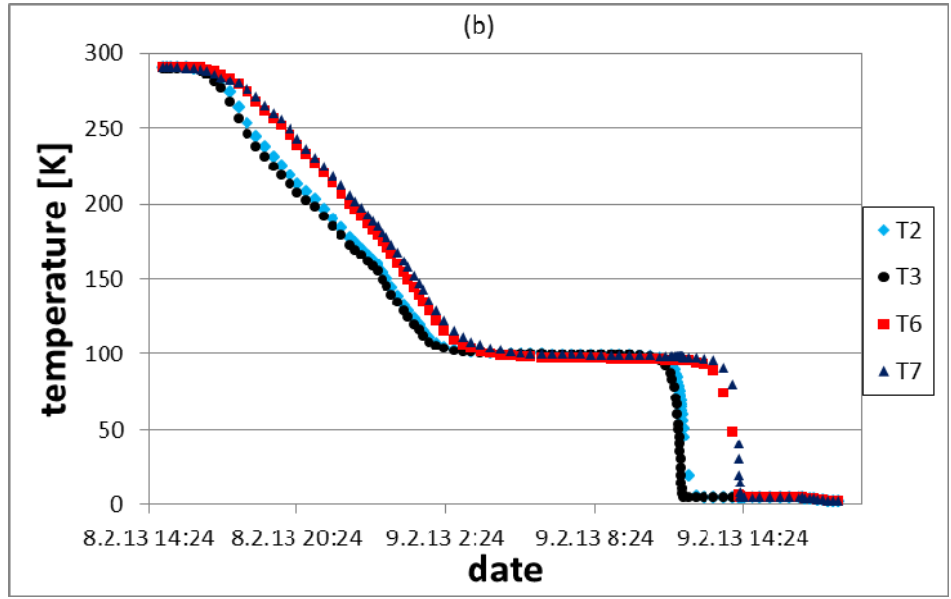

**FIGURE 6.** (a) Temperature profile during cool-down of sensors T1, T4, T5 and T8, (b) Temperature profile during cool-down of sensors T2, T3, T6 and T7.

To qualify the inserts and the cryostat for the XFEL qualification process, the maximal gradients were measured and compared to a reference value that was given by a previous measurement in a different cryostat. The values are given in table (1) and confirm that the insert is qualified for the process.

**TABLE (1).** Comparison of Cavity Gradients max. $E_{acc}$ with reference values measured in cryostat for single cavities.

| Cavity | Reference max. $E_{acc}$ [MV/m] | Measured max. $E_{acc}$ [MV/m] |
|---|---|---|
| AC122 | No reference | 17 |
| Z109 | 31 | 29 |
| Z137 | 27 | 25 |
| Z108 | 33 | 33 |

## CONCLUSION

The vertical inserts for the XATC cryostats were designed and tested successfully. All measured parameters for mechanical deformation, vacuum and particle free conditions and RF-measurements lay inside of the required limits, so that the approval for the XFEL qualification process could be confirmed. After additional test runs the cool-down and warm-up rates could be increased to a corresponding maximum thermal gradient of about 70K. A complete cool-down from room temperature to 4K can be achieved in about 14 hours.

## ACKNOWLEDGMENTS

This work was partial supported by the Commission of the European Communities under the 7[th] Framework Programme "Construction of New Infrastructures – Preparatory Phase", contract number 206711.